\documentclass[letter,10pt,conference]{IEEEtran}

 \usepackage[bookmarks=false]{hyperref}
\ifCLASSINFOpdf
\else
\fi
\usepackage{relsize}

\usepackage{centernot}
\usepackage{cite}
\usepackage{amsfonts}
\usepackage{mathrsfs}
\usepackage{bbm}
\usepackage{pifont}
\usepackage{amsfonts}
\usepackage{amsmath, amsthm}
\usepackage[top=.72in,bottom=.8in,left=.62in,right=.62in]{geometry}
\usepackage{tabularx}
\usepackage{listings}
\usepackage{graphicx}
\usepackage{float}
\usepackage{amssymb}
\usepackage{array}
\usepackage{fancyhdr}
\usepackage{cases}
 \usepackage{supertabular}
\usepackage{url}
\usepackage{fancyhdr}

\usepackage{psfrag}
\usepackage{color}
\usepackage{bbding}

\newcommand {\C} {{\rm I\kern-5.5pt C}}

\def\centerhack#1{\hbox to 0pt{\hss\footnotesize #1\hss}}
\def\centerhackn#1{\hbox to 0pt{\hss #1\hss}}
\def\dchack#1{\vbox to 0pt{\vss{\hbox to 0pt{\hss#1\hss}}\vss}}

\newtheorem{lem}{Lemma}
\newtheorem{thm}{Theorem}

\newtheorem*{proposition1.1}{Proposition 1.1}
\newtheorem*{proposition1.2}{Proposition 1.2}
\newtheorem*{proposition1.3}{Proposition 1.3}
\newtheorem*{proposition2.1}{Proposition 2.1}
\newtheorem*{proposition2.2}{Proposition 2.2}
 \renewcommand\baselinestretch{.975}

\begin{document}

%

\title{\fontsize{23.5}{24} \selectfont Threshold Functions in Random $s$-Intersection Graphs}






 \author{ \IEEEauthorblockN{Jun Zhao, Osman Ya\u{g}an and Virgil Gligor}
\IEEEauthorblockA{Electrical and Computer Engineering Department \\
Carnegie Mellon University \\
{\tt \{junzhao,oyagan,gligor\}@cmu.edu}}}

\maketitle \thispagestyle{plain} \pagestyle{plain}

%
%



\maketitle


 \begin{abstract}
 
 Random $s$-intersection graphs have recently received considerable attention in a wide range of application areas. In such a graph, 
each vertex is equipped with a set
of items in some random manner, and any two vertices establish an undirected edge in between if and only if they
have at least $s$ common items. In particular, in a \emph{uniform} random
$s$-intersection graph, each vertex \emph{independently} selects a fixed number of items uniformly at random from a common item pool, while in a \emph{binomial} random
$s$-intersection graph, each item in some item pool is \emph{independently} attached to each vertex with the same probability. 
%
%
 
 For binomial/uniform random
$s$-intersection graphs, we establish threshold functions for perfect matching containment, Hamilton cycle containment, and $k$-robustness, where $k$-robustness is in the sense of Zhang and
Sundaram \cite{6425841}. We show that these threshold functions resemble those of classical Erd\H{o}s--R\'{e}nyi graphs, where each pair of vertices has an undirected edge independently with the same probability.

\end{abstract}
%
%

 
 \section{Introduction}

\emph{Random $s$-intersection graphs} have received much interest 
recently
\cite{Rybarczyk,Perfectmatchings,Bloznelis201494,ANALCO,r1,herdingRKG,PES:6114960,EfthymiouaHM,NikoletseasHM,ryb3,zz,2013arXiv1301.0466R,yagan,ZhaoCDC}. In such a graph, 
each vertex is equipped with a set
of items in some \emph{random} manner, and two vertices establish an undirected edge in between if and only if they
share at least $s$ items. 
Random $s$-intersection graphs have been used in various applications including secure sensor networks \cite{Rybarczyk,yagan,ZhaoCDC}, social networks \cite{PES:6114960,ANALCO}, clustering \cite{PES:6114960}, and cryptanalysis \cite{herdingRKG}.

 Among different models of random $s$-intersection graphs, two widely studied models are the so-called \emph{uniform random $s$-intersection graph} and \emph{binomial random $s$-intersection graph}  \cite{Rybarczyk,ANALCO}, which are defined in detail below.



A \emph{binomial $s$-intersection graph} denoted by $G_s(n,t_n,P_n)$ is defined on $n$ vertices as follows  \cite{Rybarczyk,ANALCO}. Each
item from a pool of $P_n$ distinct
items is assigned to each vertex \emph{independently} with probability $t_n$. 
Two vertices establish an undirected edge in between
 if and only if they have no less than $s$ items in common. The word ``binomial'' is used since the number of items on each vertex follows a binomial distribution with parameters $P_n$ (the number of trials) and $t_n$ (the success probability in each trial). $t_n$ and $P_n$ are both functions of $n$, while $s$ does not scale with $n$. Also it holds that $1\leq s  \leq P_n$.


A \emph{uniform $s$-intersection graph} denoted by $H_s(n,K_n,P_n)$ is defined on $n$ vertices as follows  \cite{Rybarczyk,ANALCO}. Each vertex \emph{independently} selects $K_n$
different items \emph{uniformly at random} from a pool of $P_n$ distinct
items. Two vertices have an undirected edge in between
 if and only if they have at least $s$ common items.
 The notion ``uniform'' means that all vertices have the same number of items (but likely different sets of items). $K_n$ and $P_n$ are both functions of $n$, while $s$ does not scale with $n$. It holds that $1\leq s\leq K_n \leq P_n$.

 An important application of uniform $s$-intersection graphs is to model the topologies of secure wireless sensor networks employing
 the 
 Chan--Perrig--Song key predistribution scheme \cite{adrian}, which is widely recognized as an appropriate solution
 to secure
communications between sensors. In the  Chan--Perrig--Song key predistribution scheme for an $n$-size sensor network, 
prior to deployment,
 each sensor is assigned a set of $K_n$ distinct cryptographic keys selected uniformly at random from the same key pool containing $P_n$ different keys. After deployment, two sensors establish secure communication if
and only if they have at least $s$ common key(s). Clearly the induced topology is a uniform $s$-intersection graph.


Our main goal in this paper is to derive the threshold functions of uniform $s$-intersection graphs and binomial $s$-intersection graphs for properties including perfect matching containment, Hamilton cycle containment, and $k$-robustness. These properties are defined as follows: (i) 
A perfect matching is a set of edges that do not have common vertices
and cover all vertices with the exception of missing at most one vertex. (ii) A Hamiltonian cycle means a closed loop that visits each vertex exactly once. (iii)
The notion of $k$-robustness proposed by Zhang and
Sundaram \cite{6425841} measures the effectiveness of local-information-based diffusion algorithms in the presence of adversarial vertices; formally, a graph with a vertex set $\mathcal {V}$ is $k$-robust
 if at least one of (a) and (b) below holds for each
non-empty and strict subset $T$ of $\mathcal {V}$: (a) there exists
at least a vertex $v_a \in T$ such that
  $v_a$ has no less than $k$ neighbors inside $\mathcal {V}\setminus
  T$, and (b) there exists at least a vertex $v_b \in \mathcal {V}\setminus T$ such that
  $v_b$ has no less than $k$ neighbors inside $T$, where two vertices are neighbors if they have an edge in between.
  
  The above studied properties of uniform $s$-intersection graphs and binomial $s$-intersection graphs have diverse applications. First, in the use of uniform $s$-intersection graphs for secure wireless sensor networks \cite{adrian,Rybarczyk}, perfect matchings have been used for the optimal allocation of rate and power \cite{4407772}, the design of routing
schemes supporting data fusion \cite{4395109}, and
  the dispatch of sensors \cite{4359020} (i.e., moving sensors to areas of interest), while Hamilton cycles have been used for cyclic routing which with distributed optimization achieves efficient in-network data processing \cite{1413472}. Second, in the application of binomial $s$-intersection graphs to classification and clustering \cite{britton2008}, perfect matchings have been used to analyze 
  linear inverse problems \cite{MeshiThesis}, while Hamilton cycles have been used to study probabilistic graphical models \cite{Mengshoel20061137}. Third, the property of $k$-robustness plays
  a key role in many classes of
 dynamics in graphs, such as resilient consensus, contagion and bootstrap percolation \cite{6425841}.
  
  

  We obtain threshold functions of binomial $s$-intersection graphs and uniform $s$-intersection graphs for perfect matching containment, Hamilton cycle containment, and $k$-robustness, and show that these thresholds resemble those of Erd\H{o}s--R\'enyi graphs \cite{erdosPF}, where an Erd\H{o}s--R\'enyi graph is constructed by assigning an edge between each pair of vertices independently with the same probability. Specifically, just like Erd\H{o}s--R\'enyi graphs, for both binomial $s$-intersection graphs and uniform $s$-intersection graphs, the thresholds of the edge probability (i.e., the probability of an edge existence between two vertices) \vspace{-1pt} are given by
  \begin{itemize}
\item $\frac{ \ln n}{n}$ for perfect matching containment, \vspace{2pt}
\item $\frac{ \ln n + \ln \ln n}{n}$ for Hamilton cycle containment, and \vspace{2pt}
\item $\frac{ \ln n + (k-1) \ln \ln n}{n}$ for $k$-robustness. \vspace{-1pt}
\end{itemize}




We organize the rest of the paper as follows.  In Section \ref{sec:main:res}, we present the results as
theorems, which are proved in 
Section \ref{sec:thmprf:kcon}.
We discuss related work in  Section
\ref{related} and conclude the paper in Section \ref{sec:Conclusion}. The Appendix
provides useful lemmas and their proofs.  
  


\section{Results} \label{sec:main:res}

In Sections \ref{sect1} and \ref{sect2} below, we summarize our results of binomial random $s$-intersection graphs and uniform random $s$-intersection graphs, respectively. Afterwards, we discuss the threshold functions in Section \ref{sect3}.

Notation and convention: We denote the edge probability of a binomial random $s$-intersection graph $G_s(n,t_n,P_n)$ by $b_n$, and denote the edge probability of a binomial random $s$-intersection graph $H_s(n,K_n,P_n)$ by $u_n$. Both $k$ and $s$ are constants and do not scale with $n$.  All asymptotic statements are understood with $ \to \infty$. We use the  Landau asymptotic notation $O(\cdot), o(\cdot), \Omega(\cdot),
\omega(\cdot), \Theta(\cdot), \sim$; in particular, for two
positive sequences $x_n$ and $y_n$, the relation $x_n \sim y_n$
signifies $\lim_{n \to
  \infty} (x_n/y_n)=1$. Also, $\mathbb{P}[\mathcal {E}]$
denotes the probability that event $\mathcal {E}$ occurs. An event happens \emph{asymptotically almost surely} if its probability converges to $1$ as $n\to\infty$.

\subsection{Results of binomial random $s$-intersection graphs} \label{sect1}

We present results of a binomial random $s$-intersection graph $G_s(n,t_n,P_n)$ in Theorems 1--3 below. The conditions can be either about the edge probability $b_n$ or its  \vspace{1pt} asymptotics $\frac{1}{s!} \cdot {t_n}^{2s}{P_n}^{s} $ (our work \cite[Lemma 12]{ANALCO} proves $b_n \sim \frac{1}{s!} \cdot {t_n}^{2s}{P_n}^{s} $ under certain conditions).    \vspace{1pt}


\begin{thm}[\textbf{Perfect matching containment in binomial random $s$-intersection graphs}]  \label{pm:thm:bin}
For a binomial random $s$-intersection graph $G_s(n,t_n,P_n)$ under $P_n = \Omega(n^c)$ for
some constant $c>2-\frac{1}{s}$, under \textbf{either} of the following two conditions for all $n$ with a sequence $\alpha_n$ satisfying $\lim_{n \to \infty}{\alpha_n} = \alpha^* \in [-\infty, \infty]$:\vspace{1pt}\\
\indent (i) the edge probability $b_n$ equals 
 $ \frac{\ln  n   +
 {\alpha_n}}{n}$,  \vspace{1pt}\\
\indent (ii) $\frac{1}{s!} \cdot {t_n}^{2s}{P_n}^{s} = \frac{\ln  n   +
 {\alpha_n}}{n}$,\\
 then
\begin{align}
 \hspace{-1.27pt}\lim\limits_{n \to \infty}\hspace{-1.27pt} \mathbb{P}[\hspace{.3pt} G_s(n,t_n,P_n) \text{ contains a perfect matching.} \hspace{.3pt} ]& \hspace{-1.27pt}=\hspace{-1.27pt} e^{- e^{-\alpha^*}}\hspace{-1pt}, \nonumber
\end{align}
which implies that 
$G_s(n,t_n,P_n)$ asymptotically almost surely does not have a perfect matching if $\alpha^* = -\infty$, and asymptotically almost surely has a perfect matching if $\alpha^* = \infty$.
\end{thm}



\begin{thm}[\textbf{Hamilton cycle containment in binomial random $s$-intersection graphs}]  \label{hc:thm:bin}
For a binomial random $s$-intersection graph $G_s(n,t_n,P_n)$ under $P_n = \Omega(n^c)$ for
some constant $c>2-\frac{1}{s}$, under \textbf{either} of the following two conditions for all $n$ with a sequence $\beta_n$ satisfying $\lim_{n \to \infty}{\beta_n} = \beta^* \in [-\infty, \infty]$:\vspace{1pt}\\
\indent (i) the edge probability $b_n$ equals 
 $   \frac{\ln  n + \ln \ln n + {\beta_n}}{n}$,\vspace{1pt}  \\
\indent (ii) $\frac{1}{s!} \cdot {t_n}^{2s}{P_n}^{s} = \frac{\ln  n + \ln \ln n + {\beta_n}}{n}$,\\
 then
\begin{align}
 \hspace{-1pt}\lim\limits_{n \to \infty}\hspace{-1pt} \mathbb{P}[\hspace{.5pt} G_s(n,t_n,P_n) \text{ contains a Hamilton cycle.} \hspace{.5pt} ]& \hspace{-1pt}=\hspace{-1pt} e^{- e^{-\beta^*}}\hspace{-1pt}, \nonumber
\end{align}
which implies that 
$G_s(n,t_n,P_n)$ asymptotically almost surely does not have a Hamilton cycle if $\beta^* = -\infty$, and asymptotically almost surely has a Hamilton cycle if $\beta^* = \infty$.
\end{thm}


\begin{thm}[\textbf{$k$-Robustness in binomial random $s$-intersection graphs}]  \label{rb:thm:bin}
For a binomial random $s$-intersection graph $G_s(n,t_n,P_n)$ under $P_n = \Omega(n^c)$ for
some constant $c>2-\frac{1}{s}$, under \textbf{either} of the following two conditions for all $n$ with a sequence $\gamma_n$ satisfying $\lim_{n \to \infty}{\gamma_n} = \gamma^* \in [-\infty, \infty]$:\vspace{1pt}\\
\indent (i) the edge probability $b_n$ equals 
 $ \frac{\ln  n + {(k-1)} \ln \ln n + {\gamma_n}}{n}$,\vspace{1pt}  \\
\indent (ii) $\frac{1}{s!} \cdot {t_n}^{2s}{P_n}^{s} = \frac{\ln  n + {(k-1)} \ln \ln n + {\gamma_n}}{n}$,\\
 then
\begin{subnumcases}{ \hspace{-10pt}\lim\limits_{n \to \infty}\hspace{-1.5pt} \mathbb{P}[ G_s(n,t_n,P_n) \text{ is $k$-robust.}  ] \hspace{-1.5pt}=\hspace{-1.5pt}} 
\hspace{-3pt} 0, &\text{\hspace{-10pt}if $\gamma^*\hspace{-1.5pt}=\hspace{-1.5pt}-\infty$},  \label{RB-leq}
 \\  \hspace{-3pt}1,& \text{\hspace{-10pt}if $\gamma^*
\hspace{-1.5pt}=\hspace{-1.5pt}\infty$.}   \label{RB-geq}
 \end{subnumcases}
\end{thm}

\subsection{Results of uniform random $s$-intersection graphs} \label{sect2}

We present results of a uniform random $s$-intersection graph $H_s(n,\hspace{-.5pt}K_n,\hspace{-.5pt}P_n)$ in Theorems 4--6 below. The conditions can be either about the edge probability $u_n$ or its asymptotics $\frac{1}{s!} \cdot \frac{{K_n}^{2s}}{{P_n}^{s}} $ (our work \cite[Lemma 8]{ANALCO} shows $u_n \sim \frac{1}{s!} \cdot \frac{{K_n}^{2s}}{{P_n}^{s}} $ under certain conditions).


\begin{thm}[\textbf{Perfect matching containment in uniform random $s$-intersection graphs}]  \label{pm:thm:uni}
For a uniform random $s$-intersection graph $H_s(n,\hspace{-.5pt}K_n,\hspace{-.5pt}P_n)$ under $P_n \hspace{-.5pt}=\hspace{-.5pt} \Omega(n^c)$ for
some constant $c>2-\frac{1}{s}$, 
under \textbf{either} of the following two conditions for all $n$ with a sequence $\alpha_n$ satisfying $\lim_{n \to \infty}{\alpha_n} = \alpha^* \in [-\infty, \infty]$:\vspace{1pt}\\
\indent (i) the edge probability $u_n$ equals 
 $ \frac{\ln  n   +
 {\alpha_n}}{n}$,  \vspace{1pt}\\
\indent (ii) $\frac{1}{s!} \cdot \frac{{K_n}^{2s}}{{P_n}^{s}} = \frac{\ln  n   +
 {\alpha_n}}{n}$,\\
 then
\begin{align}
 \hspace{-1.5pt}\lim\limits_{n \to \infty}\hspace{-1.5pt} \mathbb{P}[\hspace{.3pt} H_s(n,K_n,P_n) \text{ contains a perfect matching.} \hspace{.3pt} ]& \hspace{-1.5pt}=\hspace{-1.5pt} e^{- e^{-\alpha^*}}\hspace{-1.5pt}, \label{eqx1}
\end{align}
which implies that 
$H_s(n,K_n,P_n)$ asymptotically almost surely does not have a perfect matching if $\alpha^* = -\infty$, and asymptotically almost surely has a perfect matching if $\alpha^* = \infty$.
\end{thm}


\begin{thm}[\textbf{Hamilton cycle containment in uniform random $s$-intersection graphs}]  \label{hc:thm:uni}
For a uniform random $s$-intersection graph $H_s(n,\hspace{-.5pt}K_n,\hspace{-.5pt}P_n)$ under $P_n \hspace{-.5pt}=\hspace{-.5pt} \Omega(n^c)$ for
some constant $c>2-\frac{1}{s}$, under \textbf{either} of the following two conditions for all $n$ with a sequence $\beta_n$ satisfying $\lim_{n \to \infty}{\beta_n} = \beta^* \in [-\infty, \infty]$:\vspace{1pt}\\
\indent (i) the edge probability $u_n$ equals 
 $   \frac{\ln  n + \ln \ln n + {\beta_n}}{n}$,\vspace{1pt}  \\
\indent (ii) $\frac{1}{s!} \cdot \frac{{K_n}^{2s}}{{P_n}^{s}}= \frac{\ln  n + \ln \ln n + {\beta_n}}{n}$,\\
 then
\begin{align}
 \hspace{-1pt}\lim\limits_{n \to \infty}\hspace{-1pt} \mathbb{P}[\hspace{.5pt} H_s(n,K_n,P_n) \text{ contains a Hamilton cycle.} \hspace{.5pt} ]& \hspace{-1pt}=\hspace{-1pt} e^{- e^{-\beta^*}}\hspace{-1pt},  \label{eqx2}
\end{align} 
which implies that 
$H_s(n,K_n,P_n)$ asymptotically almost surely does not have a Hamilton cycle if $\beta^* = -\infty$, and asymptotically almost surely has a Hamilton cycle if $\beta^* = \infty$.
\end{thm}

\begin{thm}[\textbf{$k$-Robustness in uniform random $s$-intersection graphs}]  \label{rb:thm:uni}
For a uniform random $s$-intersection graph $H_s(n,\hspace{-.5pt}K_n,\hspace{-.5pt}P_n)$ under $P_n \hspace{-.5pt}=\hspace{-.5pt} \Omega(n^c)$ for
some constant $c>2-\frac{1}{s}$, under \textbf{either} of the following two conditions for all $n$ with a sequence $\gamma_n$ satisfying $\lim_{n \to \infty}{\gamma_n} = \gamma^* \in [-\infty, \infty]$:\vspace{1pt}\\
\indent (i) the edge probability $u_n$ equals 
 $ \frac{\ln  n + {(k-1)} \ln \ln n + {\gamma_n}}{n}$,\vspace{1pt}  \\
\indent (ii) $\frac{1}{s!} \cdot \frac{{K_n}^{2s}}{{P_n}^{s}}= \frac{\ln  n + {(k-1)} \ln \ln n + {\gamma_n}}{n}$,\\
 then
 \begin{align}
 \hspace{-1pt}\lim\limits_{n \to \infty}\hspace{-1pt} \mathbb{P}[\hspace{.5pt} H_s(n,K_n,P_n) \text{ is $k$-robust.} \hspace{.5pt} ]& \hspace{-1pt}=\hspace{-1pt} \begin{cases} 0, &\text{ if $\gamma^*=-\infty$}, \\  1, &\text{ if $\gamma^*
=\infty$.} \end{cases}  \label{eqx3}
\end{align}
\end{thm}

\subsection{Threshold functions in random $s$-intersection graphs}   \label{sect3}

From Theorems 1--6 above and Appendix-B on Erd\H{o}s--R\'enyi graphs, we obtain that the threshold functions of binomial $s$-intersection graphs and uniform $s$-intersection graphs for the three studied properties have the same form as those of Erd\H{o}s--R\'enyi graphs. Specifically, for a binomial $s$-intersection graph, a uniform $s$-intersection graph, and an Erd\H{o}s--R\'enyi graph, 
the thresholds of the edge probability are $\frac{ \ln n}{n}$ for perfect matching containment, $\frac{ \ln n + \ln \ln n}{n}$ for Hamilton cycle containment, and $\frac{ \ln n + (k-1) \ln \ln n}{n}$ for $k$-robustness. 
%
%
%
%


 \section{Establishing Theorems \ref{pm:thm:bin}--\ref{rb:thm:uni}}
\label{sec:thmprf:kcon}

We use $\texttt{PM}$ and $\texttt{HC}$ and to stand for perfect matching and Hamilton cycle, respectively.

%
%
%
%
%
%

 \subsection{Proof of Theorem \ref{pm:thm:bin}}
 
 Theorem \ref{pm:thm:bin} follows once we prove  
\begin{align}
 \hspace{-1pt} \mathbb{P}[\hspace{.5pt} G_s(n,\hspace{-.5pt}t_n,\hspace{-.5pt}P_n) \text{ has a \texttt{PM}.} \hspace{.5pt} ]& \hspace{-1pt}\leq \hspace{-1pt} e^{- e^{-\alpha^*}}  \hspace{-1pt} \cdot \hspace{-1pt} [1\hspace{-1pt}+\hspace{-1pt}o(1)] \label{PM-leq}
 \end{align}
 and
  \begin{align}
 \hspace{-1pt} \mathbb{P}[\hspace{.5pt}G_s(n,\hspace{-.5pt}t_n,\hspace{-.5pt}P_n)\text{ has a \texttt{PM}.}\hspace{.5pt}] & \hspace{-1pt} \geq  \hspace{-1pt} e^{- e^{-\alpha^*}}  \hspace{-1pt} \cdot \hspace{-1pt} [1\hspace{-1pt}-\hspace{-1pt}o(1)]. \label{PM-geq}
 \end{align}

 (\ref{PM-leq}) clearly holds from Lemma \ref{kcon:lem:bin} in Appendix-B with $k=1$ and the fact \cite{zz} that a necessary condition for a graph to contain a \texttt{PM} is that the minimum degree is at least $1$ (i.e., there is no isolated vertex).
 
  Now we establish (\ref{PM-geq}). From Lemmas \ref{graph_Gs_cpl_KnPn} and \ref{graph_Gs_cpl_edgeprob}  in Appendix-A and the fact that \texttt{PM} containment is a monotone increasing graph property, we can introduce an auxiliary condition $|\alpha_n |= O(\ln \ln n)$. Then we explain that under $|\alpha_n |= O(\ln \ln n)$, either of conditions (i) and (ii) in Theorem \ref{pm:thm:bin} yields
   \begin{align}
\textstyle{\big|\frac{1}{s!} \cdot {t_n}^{2s}{P_n}^{s}  -  \frac{\ln  n   +
 {\alpha_n}}{n} \big|  = o\big(\frac{1}{n}\big)}. \label{KnPnnlnn}
\end{align}
  Clearly, (\ref{KnPnnlnn}) holds under condition (ii). To show (\ref{KnPnnlnn}) under condition (i) with $|\alpha_n |= O(\ln \ln n)$, we use \cite[Lemma 12]{ANALCO} to derive $\frac{1}{s!} \cdot {t_n}^{2s}{P_n}^{s} = u_n \pm o\big(\frac{1}{n}\big) = \frac{\ln  n   +
 {\alpha_n} \pm o(1)}{n} $, \vspace{1pt} which implies (\ref{KnPnnlnn}). Therefore, (\ref{KnPnnlnn}) follows, which with $|\alpha_n |= O(\ln \ln n)$ further induces
    \begin{align}
\textstyle{\frac{1}{s!} \cdot {t_n}^{2s}{P_n}^{s} = \frac{\ln  n}{n} \cdot [ 1\pm o(1)].} \label{KnPnnlnneq2}
\end{align}

  We now use Lemmas \ref{lem:ER:PM} and \ref{cp_rig_er} in the Appendix to prove (\ref{PM-geq}).  
  We show that the conditions of Lemma \ref{cp_rig_er} all hold given (\ref{KnPnnlnneq2}) and the condition on $P_n$ in Theorem \ref{pm:thm:bin}: $P_n  = \Omega(n^c)$ for
some constant $c>2-\frac{1}{s}$. We have ${t_n}^2 P_n =  \sqrt[s]{s! \cdot \big( \frac{1}{s!} \cdot {t_n}^{2s}{P_n}^{s}\big)} = \Theta \big(n^{-\frac{1}{s}} (\ln n)^{\frac{1}{s}} \big) $ \vspace{2pt} so that ${t_n}^2 P_n = o\big(\frac{1}{\ln n}\big)$ and ${t_n}^2 P_n = \omega\big( \frac{1}{n^2} \big)$.
Also, we obtain $$\textstyle{t_n  \hspace{-1.5pt}= \hspace{-3pt}  \sqrt[2s]{s!   \big( \frac{1}{s!}   {t_n}^{2s}{P_n}^{s}\big) \big/ \big({P_n}^{s} \big)}  \hspace{-1.5pt}= \hspace{-1.5pt}   O\big(\hspace{-0.5pt} (\ln  n)^{\frac{1}{2s}} n^{-\frac{1}{2}(c+\frac{1}{s})}\hspace{-0.5pt} \big)    \hspace{-1.5pt}= \hspace{-1.5pt}  o\big(\hspace{-0.5pt} \frac{1}{n} \hspace{-0.5pt}\big)  }$$
and $$\textstyle{t_n P_n   \hspace{-1.5pt}= \hspace{-3pt}  \sqrt[2s]{s!   \big( \frac{1}{s!}   {t_n}^{2s}{P_n}^{s}\big)   {P_n}^{s}}  \hspace{-1.5pt}= \hspace{-1.5pt}  \Omega(n^{\frac{cs-1}{2s}}(\ln n)^{\frac{1}{2s}})    \hspace{-1.5pt}= \hspace{-1.5pt}  \omega(\ln n),}$$
 where the last step applies $cs > 2s-1 \geq 1$. Hence, all conditions of Lemma \ref{cp_rig_er} hold. Then from Lemma \ref{mono-gcp}, Lemma \ref{cp_rig_er},  and the monotonicity of \texttt{PM} containment, there exists  a sequence $h_n$ satisfying
\begin{align}
\textstyle{h_n  = \frac{1}{s!} \cdot {t_n}^{2s}{P_n}^{s} \cdot   \big[1- o\big( \frac{1}{\ln n} \big)\big]}
\label{er-hn-prob}
\end{align}
 such that 
\begin{align}
 \mathbb{P}[ G_s(n,\hspace{-.5pt}t_n,\hspace{-.5pt}P_n) \text{ has a \texttt{PM}.}   ]  \hspace{-2pt}\geq\hspace{-2pt} \mathbb{P}[ G_{ER}(n,h_n) \text{ has a \texttt{PM}.} ] \hspace{-2pt} - \hspace{-2pt} o(1) . \label{pf-rig-er-rel}
 \end{align}

 Substituting  (\ref{KnPnnlnn}) and (\ref{KnPnnlnneq2}) into (\ref{er-hn-prob}), we derive $h_n =  \frac{\ln  n   +
 {\alpha_n} \pm o(1)}{n}$, which is used in Lemma \ref{lem:ER:PM} to induce
 \begin{align}
 \lim_{n \to \infty}   \mathbb{P}[ G_{ER}(n,h_n)\text{ has a \texttt{PM}.} ] &  = e^{- e^{-\alpha^*}} . \label{KnPnnlnn3}
 \end{align}
 Then (\ref{PM-geq}) clearly follows from (\ref{pf-rig-er-rel}) and (\ref{KnPnnlnn3}).
 
We have established  Theorem \ref{pm:thm:bin} by showing (\ref{PM-leq}) and (\ref{PM-geq}).
 
      \subsection{Proof of Theorem \ref{hc:thm:bin}}

      Theorem \ref{hc:thm:bin} follows once we prove \begin{align}
 \hspace{-1pt} \mathbb{P}[\hspace{.5pt} G_s(n,\hspace{-.5pt}t_n,\hspace{-.5pt}P_n) \text{ has a \texttt{HC}.} \hspace{.5pt} ]& \hspace{-1pt}\leq \hspace{-1pt} e^{- e^{-\beta^*}}  \hspace{-1pt} \cdot \hspace{-1pt} [1\hspace{-1pt}+\hspace{-1pt}o(1)] \label{HC-leq}
 \end{align}
 and
  \begin{align}
 \hspace{-1pt} \mathbb{P}[\hspace{.5pt}G_s(n,\hspace{-.5pt}t_n,\hspace{-.5pt}P_n)\text{ has a \texttt{HC}.}\hspace{.5pt}] & \hspace{-1pt} \geq  \hspace{-1pt} e^{- e^{-\beta^*}}  \hspace{-1pt} \cdot \hspace{-1pt} [1\hspace{-1pt}-\hspace{-1pt}o(1)]. \label{HC-geq}
 \end{align}

 (\ref{HC-leq}) clearly holds from Lemma \ref{kcon:lem:bin} with $k=2$ and the fact  \cite{zz} that a necessary condition for a graph to contain a \texttt{HC} is that the minimum degree is at least $2$.
 
  Now we establish (\ref{HC-geq}). From Lemma \ref{graph_Gs_cpl_KnPn}, Lemma \ref{graph_Gs_cpl_edgeprob} and the fact that \texttt{HC} containment is a monotone increasing graph property, we can introduce an auxiliary condition $|\beta_n |= O(\ln \ln n)$. Then we explain that under $|\beta_n |= O(\ln \ln n)$, either of conditions (i) and (ii) in Theorem \ref{hc:thm:bin} yields
   \begin{align}
 \textstyle{\big|\frac{1}{s!} \cdot {t_n}^{2s}{P_n}^{s}  -  \frac{\ln  n   + \ln \ln n +
 {\beta_n}}{n} \big|  = o\big(\frac{1}{n}\big)}. \label{KnPnnlnn-HC}
\end{align}
  Clearly, (\ref{KnPnnlnn-HC}) holds under condition (ii). To show (\ref{KnPnnlnn-HC}) under condition (i) with $|\beta_n |= O(\ln \ln n)$, we use \cite[Lemma 12]{ANALCO} to derive \vspace{1pt} $\frac{1}{s!} \cdot {t_n}^{2s}{P_n}^{s} = u_n \pm o\big(\frac{1}{n}\big) = \frac{\ln  n + \ln \ln n   +
 {\beta_n} \pm o(1)}{n} $, which implies (\ref{KnPnnlnn-HC}). Therefore, (\ref{KnPnnlnn-HC}) follows, which with $|\beta_n |= O(\ln \ln n)$ further induces (\ref{KnPnnlnneq2}). As explained above in the proof of Theorem \ref{pm:thm:bin}, all conditions of Lemma \ref{cp_rig_er} hold given (\ref{KnPnnlnneq2}) and the condition on $P_n$ in Theorem \ref{hc:thm:bin}: $P_n  = \Omega(n^c)$ for
some constant $c>2-\frac{1}{s}$.  \vspace{1pt} Then from Lemma \ref{cp_rig_er}, Lemma \ref{mono-gcp} and the monotonicity of \texttt{HC} containment, there exists  a sequence $h_n$ satisfying (\ref{er-hn-prob})
 such that
\begin{align}
 & \mathbb{P}[  G_s(n,\hspace{-.5pt}t_n,\hspace{-.5pt}P_n) \text{ has a \texttt{HC}.}  ] \hspace{-2pt}  \geq \hspace{-2pt} \mathbb{P}[ G_{ER}(n,h_n) \text{ has a \texttt{HC}.}  ] \hspace{-2pt} -\hspace{-2pt}  o(1) . \label{pf-rig-er-rel-HC}
 \end{align}
  
 Substituting  (\ref{KnPnnlnn-HC}) and (\ref{KnPnnlnneq2}) into (\ref{er-hn-prob}), we derive $h_n =  \frac{\ln  n +\ln \ln n   +
 {\beta_n} \pm o(1)}{n}$, which is used in Lemma \ref{lem:ER:HC} to induce
 \begin{align}
 \lim_{n \to \infty}   \mathbb{P}[ G_{ER}(n,h_n)\text{ has a \texttt{HC}.} ] &  = e^{- e^{-\beta^*}} . \label{KnPnnlnn-HC3}
 \end{align}
 Then (\ref{HC-geq}) clearly follows from (\ref{pf-rig-er-rel-HC}) and (\ref{KnPnnlnn-HC3}).
 
We have established  Theorem \ref{hc:thm:bin} by showing (\ref{HC-leq}) and (\ref{HC-geq}).

      \subsection{Proof of Theorem \ref{rb:thm:bin}}

%
 
From \cite[Lemma 1]{6425841}, a necessary condition for a graph to be $k$-robust is that the graph is $k$-connected, so we clearly obtain 
 (\ref{RB-leq})  from Lemma \ref{kcon:lem:bin} in view that 
 \begin{align}
 \lim_{n \to \infty}   \mathbb{P}[ G_s(n,\hspace{-.5pt}t_n,\hspace{-.5pt}P_n) \text{ is $k$-connected.} ] &  =0 ~~~\text{ if $\gamma^*=-\infty$,}
\nonumber
 \end{align}
 
  Now we establish (\ref{RB-geq}). From Lemma \ref{graph_Gs_cpl_KnPn}, Lemma \ref{graph_Gs_cpl_edgeprob} and the fact that \texttt{HC} containment is a monotone increasing graph property, we can introduce an auxiliary condition $|\gamma_n |= O(\ln \ln n)$. Then we explain that under $|\gamma_n |= O(\ln \ln n)$, either of conditions (i) and (ii) in Theorem \ref{rb:thm:bin} yields
   \begin{align}
  \textstyle{\big|\frac{1}{s!} \cdot {t_n}^{2s}{P_n}^{s}  -  \frac{\ln  n   + (k-1) \ln \ln n +
 {\gamma_n}}{n} \big|  = o\big(\frac{1}{n}\big)}. \label{KnPnnlnn-HC}
\end{align}
  Clearly, (\ref{KnPnnlnn-HC}) holds under condition (ii). To show (\ref{KnPnnlnn-HC}) under condition (i) with $|\gamma_n |= O(\ln \ln n)$, we use \cite[Lemma 12]{ANALCO} to derive $\frac{1}{s!} \cdot {t_n}^{2s}{P_n}^{s} = u_n \pm o\big(\frac{1}{n}\big) = \frac{\ln  n +(k-1) \ln \ln n   +
 {\gamma_n} \pm o(1)}{n} $, which implies (\ref{KnPnnlnn-HC}). Therefore, (\ref{KnPnnlnn-HC}) follows, which with $|\gamma_n |= O(\ln \ln n)$ further induces (\ref{KnPnnlnneq2}). As explained above in the proof of Theorem \ref{pm:thm:bin}, all conditions of Lemma \ref{cp_rig_er} hold given (\ref{KnPnnlnneq2}) and the condition on $P_n$ in Theorem \ref{rb:thm:bin}: $P_n  = \Omega(n^c)$ for
some constant $c>2-\frac{1}{s}$. Then from Lemma \ref{cp_rig_er}, Lemma \ref{mono-gcp} and the monotonicity of $k$-robustness, there exists  a sequence $h_n$ satisfying (\ref{er-hn-prob})
 such that
\begin{align}
 & \hspace{-1pt} \mathbb{P}[G_s\hspace{-.5pt}(n,\hspace{-.5pt}t_n,\hspace{-.6pt}P_n) \text{ \hspace{-1.5pt}is \hspace{-1.5pt}$k$-robust.}   ] \hspace{-2pt} \geq  \hspace{-2pt}\mathbb{P}[  G_{ER}\hspace{-.5pt}(n,\hspace{-.5pt}h_n) \text{ \hspace{-1.5pt}is\hspace{-1.5pt} $k$-robust.}  ]  \hspace{-2pt}- \hspace{-2pt} o(\hspace{-.5pt}1\hspace{-.5pt}) . \label{pf-rig-er-rel-HC}
 \end{align}

 Substituting  (\ref{KnPnnlnn-HC}) and (\ref{KnPnnlnneq2}) into (\ref{er-hn-prob}), we derive $h_n =  \frac{\ln  n +(k-1) \ln \ln n   +
 {\gamma_n} \pm o(1)}{n}$, which is used in Lemma \ref{lem:ER:RB} to induce
 \begin{align}
 \lim_{n \to \infty}   \mathbb{P}[ G_{ER}(n,h_n)\text{ is $k$-robust.} ] &  =1 ~~~\text{ if $\gamma^*=-\infty$} . \label{KnPnnlnn-HC3}
 \end{align}
 Then (\ref{RB-geq}) clearly follows from (\ref{pf-rig-er-rel-HC}) and (\ref{KnPnnlnn-HC3}).
 
We have established  Theorem \ref{rb:thm:bin} by showing (\ref{RB-leq}) and (\ref{RB-geq}).

 \subsection{Proof of Theorem \ref{pm:thm:uni}}
 
%
%
%
  
  From Lemma \ref{graph_Hs_cpl_KnPn}, Lemma \ref{graph_Hs_cpl_edgeprob} and the fact that \texttt{PM} containment is a monotone increasing graph property, we can introduce an auxiliary condition $|\alpha_n |= O(\ln \ln n)$. Then we explain that under $|\alpha_n |= O(\ln \ln n)$, either of conditions (i) and (ii) in Theorem \ref{pm:thm:uni} yields
   \begin{align}
    \textstyle{\big|\frac{1}{s!} \cdot \frac{{K_n}^{2s}}{{P_n}^{s}} -  \frac{\ln  n   +
 {\alpha_n}}{n} \big|  = o\big(\frac{1}{n}\big)}. \label{KnPnnlnn-uni-pm}
\end{align}
  Clearly, (\ref{KnPnnlnn-uni-pm}) holds under condition (ii). To show (\ref{KnPnnlnn-uni-pm}) under condition (i) with $|\alpha_n |= O(\ln \ln n)$, we use \cite[Lemma 8]{ANALCO} to derive $\frac{1}{s!} \cdot \frac{{K_n}^{2s}}{{P_n}^{s}}  = u_n \pm o\big(\frac{1}{n}\big) = \frac{\ln  n   +
 {\alpha_n} \pm o(1)}{n} $, which implies (\ref{KnPnnlnn-uni-pm}). Therefore, (\ref{KnPnnlnn-uni-pm}) follows, which with $|\alpha_n |= O(\ln \ln n)$ further induces
    \begin{align}
    \textstyle{\frac{1}{s!} \cdot \frac{{K_n}^{2s}}{{P_n}^{s}}  = \frac{\ln  n}{n} \cdot [ 1\pm o(1)].} \label{KnPnnlnn-uni-pmeq2}
\end{align}

From (\ref{KnPnnlnn-uni-pmeq2}) and $P_n  = \Omega(n^c)$ for
a constant $c>2-\frac{1}{s}$, it holds that
   \begin{align}
  \textstyle{K_n   =    \sqrt[2s]{s! \cdot \big( \frac{1}{s!} \cdot \frac{{K_n}^{2s}}{{P_n}^{s}} \big) \cdot {P_n}^{s}  } 
   = \Omega\big( n^{\frac{c}{2} - \frac{1}{2s}} (\ln n)^{\frac{1}{2s}} \big)} ,\label{Knboundnew}
   \end{align}
   which clearly implies $K_n = \omega\left( \ln n \right)$ so we obtain from Lemma
    \ref{cp_urig_brig}, Lemma \ref{mono-gcp} and the monotonicity of \texttt{PM} containment that
     \begin{align}
 & \mathbb{P}[\hspace{1pt} G_s(n, t_n^{-},P_n) \text{ has a \texttt{PM}.} \hspace{1pt} ] - o(1)
\nonumber \\ & \leq  \mathbb{P}[\hspace{1pt} H_s(n,K_n,P_n) \text{ has a \texttt{PM}.} \hspace{1pt} ]
\nonumber \\ & \leq  \mathbb{P}[\hspace{1pt} G_s(n,t_n^{+},P_n) \text{ has a \texttt{PM}.} \hspace{1pt} ] + o(1), \label{pfgrapheq}
 \end{align}
 where
 \begin{align}
 \textstyle{t_n^{\pm} = \frac{K_n}{P_n}
 \Big(1 \pm \sqrt{\frac{3\ln
n}{K_n }}\hspace{2pt}\Big).} \label{tnminus}
 \end{align}
 Then we get from (\ref{tnminus}) 
  that
   \begin{align}
 \textstyle{\frac{1}{s!} \cdot  \big({t_n^{\pm}}\big)^{2s}{P_n}^{s}  = \frac{1}{s!} \cdot \frac{{K_n}^{2s}}{{P_n}^{s}}  \cdot  \Big(1 \pm \sqrt{\frac{3\ln
n}{K_n }}\hspace{2pt}\Big)^{2s}}.  \label{tnminusplus}
\end{align}
Given (\ref{Knboundnew}) and constant $s$, we have
  \hspace{-1.5pt}    \begin{align} 
 \textstyle{\Big(1 \hspace{-1.5pt}  \pm \hspace{-1.5pt}  \sqrt{\frac{3\ln
n}{K_n }}\hspace{2pt}\Big)^{2s}  \hspace{-1.5pt}= \hspace{-1.5pt} 1\hspace{-1.5pt} \pm\hspace{-1.5pt} \Theta \Big(  \sqrt{\frac{\ln
n}{K_n }} \Big) 
\hspace{-2pt}=  \hspace{-2pt}1 \hspace{-2pt}\pm\hspace{-2pt} o\big( \frac{1}{\ln n} \big),}   \label{tnminusplusnsb}
\end{align}
 which along with (\ref{tnminusplus}) and  (\ref{KnPnnlnn-uni-pm}) under $|\alpha_n |= O(\ln \ln n)$ yields
    \begin{align}
 \textstyle{\frac{1}{s!} \cdot \big({t_n^{\pm}}\big)^{2s}{P_n}^{s}  = \frac{\ln  n   +
 {\alpha_n} \pm o(1)}{n} .}  \label{tnminusplusnb}
\end{align}
 Given (\ref{tnminusplusnb}) and $P_n  = \Omega(n^c)$ for
a constant $c>2-\frac{1}{s}$, we use Theorem \ref{pm:thm:bin} to derive
     \begin{align}
 \lim\limits_{n \to \infty} \mathbb{P}[\hspace{1pt} G_s(n, t_n^{\pm}, P_n) \text{ has a \texttt{PM}.} \hspace{1pt} ]    &  = e^{- e^{-\alpha^*}}, \nonumber
 \end{align}
 which together with (\ref{pfgrapheq}) induces (\ref{eqx1}).

\subsection{Proof of Theorem \ref{hc:thm:uni}}
 
%
%
%
  
  From Lemma \ref{graph_Hs_cpl_KnPn}, Lemma \ref{graph_Hs_cpl_edgeprob} and the fact that \texttt{HC} containment is a monotone increasing graph property, we can introduce an auxiliary condition $|\beta_n |= O(\ln \ln n)$. Then we explain that under $|\beta_n |= O(\ln \ln n)$, either of conditions (i) and (ii) in Theorem \ref{hc:thm:uni} yields
   \begin{align}
\textstyle{\big|\frac{1}{s!} \cdot \frac{{K_n}^{2s}}{{P_n}^{s}} -  \frac{\ln  n  + \ln \ln n +
 {\beta_n}}{n} \big|  = o\big(\frac{1}{n}\big)}. \label{KnPnnlnn-uni-hc}
\end{align}
  Clearly, (\ref{KnPnnlnn-uni-hc}) holds under condition (ii). To show (\ref{KnPnnlnn-uni-hc}) under condition (i) with $|\beta_n |= O(\ln \ln n)$, we use \cite[Lemma 8]{ANALCO} to derive $\frac{1}{s!} \cdot \frac{{K_n}^{2s}}{{P_n}^{s}}  = u_n \pm o\big(\frac{1}{n}\big) = \frac{\ln  n  + \ln \ln n +
 {\beta_n} \pm o(1)}{n} $, which implies (\ref{KnPnnlnn-uni-hc}). Therefore, (\ref{KnPnnlnn-uni-hc}) follows, which with $|\beta_n |= O(\ln \ln n)$ further induces (\ref{KnPnnlnn-uni-pmeq2}). Then (\ref{Knboundnew}) holds, and we obtain from Lemma
    \ref{cp_urig_brig}, Lemma \ref{mono-gcp} and the monotonicity of \texttt{HC} containment that
     \begin{align}
 & \mathbb{P}[\hspace{1pt} G_s(n, t_n^{-},P_n) \text{ has a \texttt{HC}.} \hspace{1pt} ] - o(1)
\nonumber \\ & \leq  \mathbb{P}[\hspace{1pt} H_s(n,K_n,P_n) \text{ has a \texttt{HC}.} \hspace{1pt} ]
\nonumber \\ & \leq  \mathbb{P}[\hspace{1pt} G_s(n,t_n^{+},P_n) \text{ has a \texttt{HC}.} \hspace{1pt} ] + o(1), \label{hcgrapheq}
 \end{align}
 with $ t_n^{-}$ and $ t_n^{+} $ specified in 
  (\ref{tnminus}). 
   Then we also obtain (\ref{tnminusplus}) and (\ref{tnminusplusnsb}), which together with   (\ref{KnPnnlnn-uni-pm}) under $|\beta_n |= O(\ln \ln n)$ lead to
      \begin{align}
 \frac{1}{s!} \cdot \big({t_n^{\pm}}\big)^{2s}{P_n}^{s} & = \frac{\ln  n   + \ln \ln n +
 {\beta_n} \pm o(1)}{n} .  \label{tnminusplusnb-HC}
\end{align}
 Given (\ref{tnminusplusnb-HC}) and $P_n  = \Omega(n^c)$ for
a constant $c>2-\frac{1}{s}$, we use Theorem \ref{hc:thm:bin} to derive
     \begin{align}
 \lim\limits_{n \to \infty} \mathbb{P}[\hspace{1pt} G_s(n, t_n^{\pm}, P_n) \text{ has a \texttt{HC}.} \hspace{1pt} ]    &  = e^{- e^{-\beta^*}}, \nonumber
 \end{align}
 which along with (\ref{hcgrapheq}) yields (\ref{eqx2}).

\subsection{Proof of Theorem \ref{rb:thm:uni}}

  From Lemma \ref{graph_Hs_cpl_KnPn}, Lemma \ref{graph_Hs_cpl_edgeprob} and the fact that $k$-robustness is a monotone increasing graph property, we can introduce an auxiliary condition $|\gamma_n |= O(\ln \ln n)$. Then we explain that under $|\gamma_n |= O(\ln \ln n)$, either of conditions (i) and (ii) in Theorem \ref{rb:thm:uni} yields
   \begin{align}
\textstyle{\big|\frac{1}{s!} \cdot \frac{{K_n}^{2s}}{{P_n}^{s}} -  \frac{\ln  n  + (k-1)\ln \ln n +
 {\gamma_n}}{n} \big|  = o\big(\frac{1}{n}\big)}. \label{KnPnnlnn-uni-rb}
\end{align}
  Clearly, (\ref{KnPnnlnn-uni-rb}) holds under condition (ii). To show (\ref{KnPnnlnn-uni-rb}) under condition (i) with $|\gamma_n |= O(\ln \ln n)$, we use \cite[Lemma 8]{ANALCO} to derive $\frac{1}{s!} \cdot \frac{{K_n}^{2s}}{{P_n}^{s}}  = u_n \pm o\big(\frac{1}{n}\big) = \frac{\ln  n   + (k-1)\ln \ln n +
 {\gamma_n} \pm o(1)}{n} $, which implies (\ref{KnPnnlnn-uni-rb}). Therefore, (\ref{KnPnnlnn-uni-rb}) follows, which with $|\gamma_n |= O(\ln \ln n)$ further induces (\ref{KnPnnlnn-uni-pmeq2}). Then (\ref{Knboundnew}) holds, and we obtain from Lemma
    \ref{cp_urig_brig}, Lemma \ref{mono-gcp} and the monotonicity of $k$-robustness that
     \begin{align}
 & \mathbb{P}[\hspace{1pt} G_s(n, t_n^{-},P_n)\text{ is $k$-robust.} \hspace{1pt} ] - o(1)
\nonumber \\ & \leq  \mathbb{P}[\hspace{1pt} H_s(n,K_n,P_n) \text{ is $k$-robust.} \hspace{1pt} ]
\nonumber \\ & \leq  \mathbb{P}[\hspace{1pt} G_s(n,t_n^{+},P_n) \text{ is $k$-robust.}  \hspace{1pt} ] + o(1), \label{rbgrapheq}
 \end{align}
 with $ t_n^{-}$ and $ t_n^{+} $ specified in 
  (\ref{tnminus}). 
   Then we also obtain (\ref{tnminusplus}) and (\ref{tnminusplusnsb}), which along with   (\ref{KnPnnlnn-uni-pm}) under $|\gamma_n |= O(\ln \ln n)$ result in
      \begin{align}
 \frac{1}{s!} \cdot \big({t_n^{\pm}}\big)^{2s}{P_n}^{s} & = \frac{\ln  n   + (k-1) \ln \ln n +
 {\gamma_n} \pm o(1)}{n} .  \label{tnminusplusnb-RB}
\end{align}
 Given (\ref{tnminusplusnb-RB}) and $P_n  = \Omega(n^c)$ for
a constant $c>2-\frac{1}{s}$, we use Theorem \ref{rb:thm:bin} to derive
     \begin{align}
 \lim\limits_{n \to \infty} \mathbb{P}[\hspace{1pt} G_s(n, t_n^{\pm}, P_n) \text{ is $k$-robust.} \hspace{1pt} ]    &  = \begin{cases} 0, &\text{ if $\gamma^*=-\infty$}, \\  1, &\text{ if $\gamma^*
=\infty$,}  \end{cases}  \nonumber
 \end{align}
 which together with (\ref{rbgrapheq}) yields (\ref{eqx3}).

\section{Related Work} \label{related}

 Binomial $s$-intersection graphs have been studied as follows.
  For $k$-connectivity, we \cite{ANALCO} obtain the asymptotically exact probability and specify $ \frac{\ln  n  + (k-1)\ln \ln n }{n} $ as a threshold of the edge probability. Bloznelis \emph{et al.} \cite{Rybarczyk} investigate the component evolution in binomial $s$-intersection graphs and prove $ \frac{1   }{n} $ as a threshold of the edge probability for the emergence of a giant component (i.e., a connected subgraph of $\Theta(n)$ vertices).

 Uniform $s$-intersection graphs have also been investigated as follows.
For perfect matching containment, Bloznelis and {\L}uczak \cite{Perfectmatchings} give the asymptotically exact probability result, which determines $ \frac{\ln  n   }{n} $ as a threshold of the edge probability,
but their result after a rewriting applies to a different set of conditions on $P_n$ compared with our Theorem \ref{pm:thm:bin}. We require $P_n  = \Omega(n^c)$ for
a constant $c>2-\frac{1}{s}$, while they \vspace{1pt} consider instead a narrow range of $P_n = \Omega\big(n (\ln n)^{-1}\big)$ and $P_n = o\big(n (\ln n)^{-\frac{3}{5s}}\big)$.
 For $k$-connectivity, both our recent paper \cite{ANALCO} and another work by Bloznelis and Rybarczyk \cite{Bloznelis201494} derive the asymptotically exact probability and determine $ \frac{\ln  n  + (k-1)\ln \ln n }{n} $ as a threshold of the edge probability. However, our result \cite{ANALCO} considers $P_n = \Omega(n)$ for $s\geq 2$ or $P_n = \Omega(n^c)$ for $s=1$ with a constant $c>1$, while Bloznelis and Rybarczyk \cite{Bloznelis201494} again \vspace{1pt} use $P_n = \Omega\big(n (\ln n)^{-1}\big)$ and $P_n = o\big(n (\ln n)^{-\frac{3}{5s}}\big)$.
 Bloznelis \emph{et al.} \cite{Rybarczyk} regard the component evolution in uniform $s$-intersection graphs and show $ \frac{1   }{n} $ as a threshold of the edge probability for the appearance of a giant component.

 A large body of work \cite{zz,2013arXiv1301.0466R,ZhaoCDC,NikoletseasHM,EfthymiouaHM,ryb3,r1,yagan} study binomial/uniform $\boldsymbol{1}$-intersection graphs as follows: 
 Rybarczyk \cite{zz,2013arXiv1301.0466R} investigates $k$-connectivity, perfect matching containment and Hamilton cycle containment; we \cite{ZhaoCDC} consider $k$-robustness and $k$-connectivity; Efthymioua and Spirakis \cite{EfthymiouaHM} and Nikoletseas \emph{et al. } \cite{NikoletseasHM} analyze Hamilton cycle containment; and Blackburn and Gerke \cite{r1}, Rybarczyk
\cite{ryb3,zz,2013arXiv1301.0466R}, and Ya\u{g}an and Makowski
\cite{yagan} look at connectivity.

\section{Conclusion}
\label{sec:Conclusion}

In this paper, for binomial/uniform random
$s$-intersection graphs, we establish threshold functions for perfect matching containment, Hamilton cycle containment, and $k$-robustness. To obtain these results, we derive the asymptotically exact probabilities of perfect matching containment and Hamilton cycle containment, and zero--one laws for $k$-robustness.




 \renewcommand\baselinestretch{.95}

 \renewcommand\baselinestretch{1}

\appendix  


\subsection{Confining $\alpha_n$ in Theorems 1 and 4, $\beta_n$ in Theorems 2 and 5, and $\gamma_n$ in Theorems 3 and 6 all as $\pm O(\ln \ln n)$}

\begin{lem}[\hspace{0pt}{Our work \cite[Lemma 2]{ANALCO}}\hspace{0pt}]  \label{graph_Gs_cpl_KnPn}

For a binomial random intersection graph
$G_s(n,t_n,P_n)$ under $P_n = \Omega(n^c)$ for
a constant $c>2-\frac{1}{s}$ and $\frac{1}{s!} \cdot {t_n}^{2s}{P_n}^{s}    = \frac{\ln  n + {(k-1)} \ln \ln n + {\alpha_n}}{n} $, the following results hold:

(i) If $\lim_{n \to \infty}\alpha_n = -\infty$, then there exists graph $G_s(n,\widetilde{t_n},\widetilde{P_n})$ under $\widetilde{P_n} = \Omega(n^{\widetilde{c}})$ with a constant $\widetilde{c}>2-\frac{1}{s}$, and 
$\frac{1}{s!} \cdot {\widetilde{t_n}}^{2s}{\widetilde{P_n}}^{s}    = \frac{\ln  n + {(k-1)} \ln \ln n + {\widetilde{\alpha_n}}}{n} $
with $\lim_{n \to \infty}\widetilde{\alpha_n} = -\infty$ and $\widetilde{\alpha_n} = -O(\ln \ln n)$,
such that $G_s(n,t_n,P_n) \preceq G_s(n,\widetilde{t_n},\widetilde{P_n})$.

(ii) If $\lim_{n \to \infty}\alpha_n = \infty$, then there exists graph $G_s(n,\widehat{t_n},\widehat{P_n})$ under $\widehat{P_n} = \Omega(n^{\widehat{c}})$ with a constant $\widehat{c}>2-\frac{1}{s}$, and 
$\frac{1}{s!} \cdot {\widehat{t_n}}^{2s}{\widehat{P_n}}^{s}     = \frac{\ln  n + {(k-1)} \ln \ln n + {\widehat{\alpha_n}}}{n}$
with $\lim_{n \to \infty}\widehat{\alpha_n} = \infty$ and $\widehat{\alpha_n} = O(\ln \ln n)$,
such that $G_s(n,\widehat{t_n},\widehat{P_n}) \preceq G_s(n,t_n,P_n)  $.

\end{lem}

\begin{lem}[\hspace{0pt}{Our work \cite[Lemma 16]{ANALCO}}\hspace{0pt}]  \label{graph_Gs_cpl_edgeprob}

For a binomial random intersection graph
$G_s(n,t_n,P_n)$ under $P_n = \Omega(n^c)$ for
a constant $c>2-\frac{1}{s}$ and $b_n   = \frac{\ln  n + {(k-1)} \ln \ln n + {\alpha_n}}{n} $, where $b_n$ is the edge probability of $G_s(n,t_n,P_n)$, the following results hold:

(i) If $\lim_{n \to \infty}\alpha_n = -\infty$, then there exists graph $G_s(n,\widetilde{t_n},\widetilde{P_n})$ under $\widetilde{P_n} = \Omega(n^{\widetilde{c}})$ with a constant $\widetilde{c}>2-\frac{1}{s}$, and 
$\widetilde{b_n}  = \frac{\ln  n + {(k-1)} \ln \ln n + {\widetilde{\alpha_n}}}{n} $
with $\lim_{n \to \infty}\widetilde{\alpha_n} = -\infty$ and $\widetilde{\alpha_n} = -O(\ln \ln n)$, where $\widetilde{b_n}$ is the edge probability of $G_s(n,\widetilde{t_n},\widetilde{P_n})$,
such that $G_s(n,t_n,P_n) \preceq G_s(n,\widetilde{t_n},\widetilde{P_n})$.

(ii) If $\lim_{n \to \infty}\alpha_n = \infty$, then there exists graph $G_s(n,\widehat{t_n},\widehat{P_n})$ under $\widehat{P_n} = \Omega(n^{\widehat{c}})$ with a constant $\widehat{c}>2-\frac{1}{s}$, and 
$\widehat{b_n}   = \frac{\ln  n + {(k-1)} \ln \ln n + {\widehat{\alpha_n}}}{n}$
with $\lim_{n \to \infty}\widehat{\alpha_n} = \infty$ and $\widehat{\alpha_n} = O(\ln \ln n)$, where $\widehat{b_n}$ is the edge probability of $G_s(n,\widehat{t_n},\widehat{P_n})$,
such that $G_s(n,\widehat{t_n},\widehat{P_n}) \preceq G_s(n,t_n,P_n)  $.

\end{lem}

\begin{lem}[\hspace{0pt}{Our work \cite[Lemma 1]{ANALCO}}\hspace{0pt}]  \label{graph_Hs_cpl_KnPn}

For a uniform random $s$-intersection graph $H_s(n,\hspace{-.5pt}K_n,\hspace{-.5pt}P_n)$ under $P_n \hspace{-.5pt}=\hspace{-.5pt} \Omega(n^c)$ for
a constant $c>2-\frac{1}{s}$ and $\frac{1}{s!} \cdot \frac{{K_n}^{2s}}{{P_n}^{s}}  = \frac{\ln  n + {(k-1)} \ln \ln n + {\beta_n}}{n} $, the following results hold:

(i) If $\lim_{n \to \infty}\beta_n = -\infty$, then there exists graph $H_s(n,\widetilde{K_n},\widetilde{P_n})$ under $\widetilde{P_n} = \Omega(n^{\widetilde{c}})$ with a constant $\widetilde{c}>2-\frac{1}{s}$, and 
$\frac{1}{s!} \cdot \frac{{\widetilde{K_n}}^{2s}}{{\widetilde{P_n}}^{s}}    = \frac{\ln  n + {(k-1)} \ln \ln n + {\widetilde{\beta_n}}}{n} $
with $\lim_{n \to \infty}\widetilde{\beta_n} = -\infty$ and $\widetilde{\beta_n} = -O(\ln \ln n)$,
such that $H_s(n,K_n,P_n) \preceq H_s(n,\widetilde{K_n},\widetilde{P_n})$.

(ii) If $\lim_{n \to \infty}\beta_n = \infty$, then there exists graph $H_s(n,\widehat{K_n},\widehat{P_n})$ under $\widehat{P_n} = \Omega(n^{\widehat{c}})$ with a constant $\widehat{c}>2-\frac{1}{s}$, and 
$\frac{1}{s!} \cdot \frac{{\widehat{K_n}}^{2s}}{{\widehat{P_n}}^{s}}    = \frac{\ln  n + {(k-1)} \ln \ln n + {\widehat{\beta_n}}}{n}$
with $\lim_{n \to \infty}\widehat{\beta_n} = \infty$ and $\widehat{\beta_n} = O(\ln \ln n)$,
such that $H_s(n,\widehat{K_n},\widehat{P_n}) \preceq H_s(n,K_n,P_n)  $.

\end{lem}

\begin{lem}[\hspace{0pt}{Our work \cite[Lemma 15]{ANALCO}}\hspace{0pt}]  \label{graph_Hs_cpl_edgeprob}

For a uniform random $s$-intersection graph $H_s(n,\hspace{-.5pt}K_n,\hspace{-.5pt}P_n)$ under $P_n \hspace{-.5pt}=\hspace{-.5pt} \Omega(n^c)$ for
a constant $c>2-\frac{1}{s}$ and $u_n  = \frac{\ln  n + {(k-1)} \ln \ln n + {\beta_n}}{n} $, where $u_n$ is the edge probability of $H_s(n,\hspace{-.5pt}K_n,\hspace{-.5pt}P_n)$, the following results hold:

(i) If $\lim_{n \to \infty}\beta_n = -\infty$, then there exists graph $H_s(n,\widetilde{K_n},\widetilde{P_n})$ under $\widetilde{P_n} = \Omega(n^{\widetilde{c}})$ with a constant $\widetilde{c}>2-\frac{1}{s}$, and 
$\widetilde{u_n}  = \frac{\ln  n + {(k-1)} \ln \ln n + {\widetilde{\beta_n}}}{n} $
with $\lim_{n \to \infty}\widetilde{\beta_n} = -\infty$ and $\widetilde{\beta_n} = -O(\ln \ln n)$, where $\widetilde{u_n}$ is the edge probability of $H_s(n,\widetilde{K_n},\widetilde{P_n})$,
such that $H_s(n,K_n,P_n) \preceq H_s(n,\widetilde{K_n},\widetilde{P_n})$.

(ii) If $\lim_{n \to \infty}\beta_n = \infty$, then there exists graph $H_s(n,\widehat{K_n},\widehat{P_n})$ under $\widehat{P_n} = \Omega(n^{\widehat{c}})$ with a constant $\widehat{c}>2-\frac{1}{s}$, and 
$\widehat{u_n}   = \frac{\ln  n + {(k-1)} \ln \ln n + {\widehat{\beta_n}}}{n}$
with $\lim_{n \to \infty}\widehat{\beta_n} = \infty$ and $\widehat{\beta_n} = O(\ln \ln n)$, where $\widehat{u_n}$ is the edge probability of $H_s(n,\widehat{K_n},\widehat{P_n})$,
such that $H_s(n,\widehat{K_n},\widehat{P_n}) \preceq H_s(n,K_n,P_n)  $.

\end{lem}

\subsection{Our previous work on random $s$-intersection graphs for $k$-connectivity and the property of minimum degree being at least $k$} 

\begin{lem}[\hspace{0pt}{Our work \cite[Theorem 2 and Lemma 14]{ANALCO}}\hspace{0pt}]   \label{kcon:lem:bin}
For a binomial random $s$-intersection graph $G_s(n,t_n,P_n)$ under $P_n = \Omega(n^c)$ for
a constant $c>2-\frac{1}{s}$, under \textbf{either} of the following two conditions for all $n$ with a sequence $\delta_n$ with $\lim_{n \to \infty}{\delta_n} = \delta^* \in [-\infty, \infty]$:\vspace{1pt}\\
\indent (i) the edge probability $b_n$ equals 
 $ \frac{\ln  n + {(k-1)} \ln \ln n + {\delta_n}}{n}$,\vspace{1pt}  \\
\indent (ii) $\frac{1}{s!} \cdot {t_n}^{2s}{P_n}^{s} = \frac{\ln  n + {(k-1)} \ln \ln n + {\delta_n}}{n}$,\\
 then
 \begin{align}
& \hspace{-1pt}\lim\limits_{n \to \infty}\hspace{-1pt} \mathbb{P}[\hspace{.5pt} G_s(n,t_n,P_n) \text{ is $k$-connected.} \hspace{.5pt} ] \hspace{-1pt} \nonumber \\ & =   \hspace{-1pt}\lim\limits_{n \to \infty}\hspace{-1pt} \mathbb{P}[\hspace{.5pt} G_s(n,t_n,P_n) \text{ has a minmimum degree at least $k$.} \hspace{.5pt} ] \nonumber \\ & =  \hspace{-1pt}e^{-
\frac{e^{-\delta ^*}}{(k-1)!}}. \nonumber
\end{align}
\end{lem}

\begin{lem}[\hspace{0pt}{Our work \cite[Theorem 1 and Lemma 13]{ANALCO}}\hspace{0pt}]  \label{kcon:lem:uni}
For a uniform random $s$-intersection graph $H_s(n,\hspace{-.5pt}K_n,\hspace{-.5pt}P_n)$ under $P_n \hspace{-.5pt}=\hspace{-.5pt} \Omega(n^c)$ for
a constant $c>2-\frac{1}{s}$, under \textbf{either} of the following two conditions for all $n$ with a sequence $\delta_n$ with $\lim_{n \to \infty}{\delta_n} = \delta^* \in [-\infty, \infty]$:\vspace{1pt}\\
\indent (i) the edge probability $u_n$ equals 
 $ \frac{\ln  n + {(k-1)} \ln \ln n + {\delta_n}}{n}$,\vspace{1pt}  \\
\indent (ii) $\frac{1}{s!} \cdot \frac{{K_n}^{2s}}{{P_n}^{s}} = \frac{\ln  n + {(k-1)} \ln \ln n + {\delta_n}}{n}$,\\
 then
 \begin{align}
& \hspace{-1pt}\lim\limits_{n \to \infty}\hspace{-1pt} \mathbb{P}[\hspace{.5pt} H_s(n,\hspace{-.5pt}K_n,\hspace{-.5pt}P_n) \text{ is $k$-connected.} \hspace{.5pt} ] \hspace{-1pt} \nonumber \\ & =   \hspace{-1pt}\lim\limits_{n \to \infty}\hspace{-1pt} \mathbb{P}[\hspace{.5pt} H_s(n,\hspace{-.5pt}K_n,\hspace{-.5pt}P_n) \text{ has a minmimum degree at least $k$.} \hspace{.5pt} ] \nonumber \\ & =  \hspace{-1pt}e^{-
\frac{e^{-\delta ^*}}{(k-1)!}}. \nonumber
\end{align}
\end{lem}

\subsection{Prior work on Erd\H{o}s--R\'enyi graphs for perfect matching containment, Hamilton cycle containment and $k$-robustness} \label{sec-er-result}

\begin{lem}[\hspace{-.1pt}{\cite[Theorem 1]{erdosPF}}\hspace{0pt}] \label{lem:ER:PM}
For an Erd\H{o}s--R\'enyi graph $G_{ER}(n,h_n)$, if there is a sequence $\alpha_n$ with $\lim_{n \to \infty}{\alpha_n} \in [-\infty, \infty]$
such that $h_n  = \frac{\ln  n   +
 {\alpha_n}}{n}$, 
 then it holds that
 \begin{align}
 \lim_{n \to \infty}   \mathbb{P}[ G_{ER}(n,h_n)\text{ has a perfect matching.} ] &  = e^{- e^{-\lim\limits_{n \to \infty}{\alpha_n}}}. \nonumber
 \end{align}
 \end{lem}

 \begin{lem} [\hspace{-.1pt}{\cite[Theorem 1]{erdosHC}}\hspace{0pt}] \label{lem:ER:HC}
For an Erd\H{o}s--R\'enyi graph $G_{ER}(n,h_n)$, if there is a sequence $\beta_n$ with $\lim_{n \to \infty}{\beta_n} \in [-\infty, \infty]$
such that $h_n = \frac{\ln  n   + \ln \ln n +
 {\beta_n}}{n}$,  
 then it holds that
 \begin{align}
 \lim_{n \to \infty}   \mathbb{P}[ G_{ER}(n,h_n)\text{ has a Hamilton cycle.} ] &  = e^{- e^{-\lim\limits_{n \to \infty}{\beta_n}}}. \nonumber
 \end{align}
 \end{lem} 
 
 \begin{lem} [\hspace{-.1pt}{Our work \cite[Lemma 1]{ZhaoCDC} based on \cite[Theorem 3]{6425841}}\hspace{0pt}]
\label{lem:ER:RB} For an Erd\H{o}s--R\'{e}nyi graph
$G(n,h_n)$, with a sequence $\gamma_n$ for all $n$ through \vspace{-2pt}
\begin{align}
h_n = & \frac{\ln  n + {(k-1)} \ln \ln n + {\gamma_n}}{n}
 \label{hat_pn},   \vspace{-2pt} 
 \end{align}
then it holds that \vspace{-2pt}
\begin{align}
 & \lim_{n \to \infty} \hspace{-1pt}\mathbb{P} \big[G(n,h_n)\textrm{
is $k$-robust}.\big]  = \begin{cases} 0, \textrm{ if $\lim_{n \to
\infty}{\gamma_n} \hspace{-2pt}=\hspace{-2pt}-\infty$}, \\  1,
\textrm{ if $\lim_{n \to \infty}{\gamma_n}
\hspace{-2pt}=\hspace{-2pt}\infty$.}
\end{cases} \vspace{-2pt} \label{er_rb}
 \end{align}
\end{lem} 
 
\subsection{A coupling between random graphs} 

Intuitively, a coupling between random graphs is used so that results on the probability of one graph having certain monotone property can help obtain the result on the probability of another graph having the same property  \cite{zz,2013arXiv1301.0466R,ZhaoCDC}. As explained by 
Rybarczyk \cite{zz,2013arXiv1301.0466R}, a coupling
of two random graphs
$G_1$ and $G_2$ means a probability space on which random graphs
$G_1'$ and $G_2'$ are defined such that $G_1'$ and $G_2'$ have the
same distributions as $G_1$ and $G_2$, respectively. If $G_1'$ is a spanning
subgraph (resp., spanning supergraph) $G_2'$, we say that under the graph coupling, $G_1$ is a spanning subgraph (resp., spanning supergraph) $G_2$. 

Following Rybarczyk's notation \cite{zz}, we
write
\begin{align}
G_1 \preceq  G_2 \quad (\textrm{resp.}, G_1 \preceq_{1-o(1)} G_2)
\label{g1g2coupling}
\end{align}
if there exists a coupling under which $G_1$ is a spanning subgraph
of $G_2$ with probability $1$ (resp., $1-o(1)$).

For two random graphs $G_1$ and $G_2$, with $\mathcal {I}$ being a monotone increasing graph property,
the following lemma relates $\mathbb{P} \big[ \hspace{1pt} G_1\text{
has $\mathcal {I}$.} \hspace{1pt}\big]$ and $\mathbb{P} \big[ \hspace{1pt} G_2\text{
has $\mathcal {I}$.} \hspace{1pt}\big]$.

the probability that

 \begin{lem}[Rybarczyk \cite{zz}]  \label{mono-gcp}

For two random graphs $G_1$ and $G_2$, the following results hold
for any monotone increasing graph property $\mathcal {I}$.

(i) If $G_1 \preceq G_2$, then 
\begin{align}
\mathbb{P} \big[ \hspace{1pt} G_2\text{
has $\mathcal {I}$.} \hspace{1pt}\big] \geq \mathbb{P} \big[ \hspace{1pt} G_1\text{ has
$\mathcal {I}$.} \hspace{1pt}\big].
 \end{align}

(ii) If $G_1 \preceq_{1-o(1)}  G_2$, then 
\begin{align}
\mathbb{P} \big[\hspace{1pt} G_2\text{
has $\mathcal {I}$.} \hspace{1pt}\big] \geq \mathbb{P} \big[ \hspace{1pt}G_1\text{ has
$\mathcal {I}$.}\hspace{1pt}\big]- o(1).
 \end{align}
 
\end{lem}

\subsection{Containment of Erd\H{o}s--R\'enyi graphs in binomial random intersection graphs} 

\begin{lem}[\hspace{0pt}{Our work \cite[Lemma 5]{ICD}}\hspace{0pt}]  \label{cp_rig_er}
If ${t_n}^2 P_n = o\big( \frac{1}{\ln n} \big)$, ${t_n}^2 P_n = \omega\big( \frac{1}{n^2} \big)$, ${t_n}=o\big( \frac{1}{n} \big)$, and $t_n  P_n = \omega(\ln n)$, then
there exits some $h_n$ satisfying
\begin{align}
h_n & = \frac{1}{s!} \cdot {t_n}^{2s}{P_n}^{s} \cdot \bigg[1- o\bigg( \frac{1}{\ln n} \bigg)\bigg]
\label{pnpb01}
\end{align}
such that an Erd\H{o}s--R\'{e}nyi graph $G_{ER}(n,h_n)$ 
 and a binomial random intersection graph $G_s(n,\hspace{-.5pt}t_n,\hspace{-.5pt}P_n)$ obey
\begin{align}
 G_{ER}(n,h_n) & \preceq_{1-o(1)} G_s(n,\hspace{-.5pt}t_n,\hspace{-.5pt}P_n). \label{GerGb}
\end{align}
 \end{lem}

\subsection{Couplings between a binomial random $s$-intersection graph an a uniform random $s$-intersection graph}

\begin{lem}[{\hspace{-0.2pt}\cite[Lemma 4]{Rybarczyk}}] \label{rkgikg} If $t_n P_n = \omega\left( \ln n \right)$, and for all $n$
sufficiently large,
\begin{align}
K_{n}^{-}  & \leq t_n P_n - \sqrt{3(t_n P_n + \ln n) \ln n} ,
\nonumber \\
K_{n}^{+}  & \geq t_n P_n + \sqrt{3(t_n P_n + \ln n) \ln n}  ,
\nonumber
\end{align}
then
\begin{align}
 H_s(n,K_{n}^{-}, P_n) &  \preceq_{1-o(1)}  G_s(n,t_n,P_n) \nonumber
\\  & \preceq_{1-o(1)}  H_s(n, K_{n}^{+}, P_n).  \nonumber  
 \end{align}
\end{lem}

\begin{lem} \label{cp_urig_brig}

If $K_n = \omega\left( \ln n \right)$, then
with $t_n^{-} = \frac{K_n}{P_n}
 \left(1 - \sqrt{\frac{3\ln
n}{K_n }}\hspace{2pt}\right)$ and $t_n^{+} = \frac{K_n}{P_n}
 \left(1 + \sqrt{\frac{3\ln
n}{K_n }}\hspace{2pt}\right)$,  
it holds that   
\begin{align}
G_s(n,t_n^{-},P_n) & \preceq_{1-o(1)}  H_s(n,K_n,P_n) \nonumber
\\  &  \preceq_{1-o(1)}  G_s(n,t_n^{+},P_n) . \nonumber
 \end{align}

\end{lem}

\subsection{The Proof of Lemma \ref{cp_urig_brig}}

We use Lemma \ref{rkgikg} to prove Lemma \ref{cp_urig_brig}. From
conditions\vspace{1pt} $K_n = \omega\left( \ln n \right)$ and $t_n^{\pm} =
\frac{K_n}{P_n}
 \left(1 \pm \sqrt{\frac{3\ln
n}{K_n }}\hspace{2pt}\right)$, it holds that $t_n^{\pm} P_n  = \omega\left( \ln n \right)$.
 For all $n$ sufficiently large, we obtain
\begin{align}
&  K_n - \left[ t_n^{-} P_n + \sqrt{3(t_n^{-} P_n + \ln n) \ln n}
\hspace{1.5pt}\right] \nonumber \\ & = K_n \sqrt{\frac{3\ln n}{K_n
}} - \sqrt{3\left[ K_n \left(1 - \sqrt{\frac{3\ln n}{K_n
}}\hspace{2pt}\right) + \ln n\right] \ln n} \nonumber
\\  & = \sqrt{3K_n\ln n}  -
\sqrt{3\left[K_n  \hspace{-1pt}+ \hspace{-1pt} \sqrt{\ln n} \left(
\sqrt{\ln n} \hspace{-1pt}- \hspace{-1pt}
\sqrt{3K_n}\hspace{2pt}\right) \right ] \hspace{-1pt} \ln n}
\nonumber \\  & \geq \sqrt{3K_n\ln n}  - \sqrt{3K_n\ln n}    \nonumber \\  & =  0 \nonumber
\end{align}
and
\begin{align}
&  K_n - \left[ t_n^{+} P_n - \sqrt{3(t_n^{+} P_n + \ln n) \ln n}
\hspace{1.5pt}\right] \nonumber \\ & =- K_n \sqrt{\frac{3\ln n}{K_n
}} + \sqrt{3\left[ K_n \left(1 + \sqrt{\frac{3\ln n}{K_n
}}\hspace{2pt}\right) + \ln n\right] \ln n} \nonumber
 \\  &  \leq - \sqrt{3K_n\ln n}  + \sqrt{3K_n\ln n}    \nonumber \\  & =  0. \nonumber
\end{align}
Then by  Lemma \ref{rkgikg}, we have $G_s(n,t_n^{-},P_n)   \preceq_{1-o(1)}  H_s(n, K_{n}, P_n)$ and $ H_s(n,K_{n}, P_n)  \preceq_{1-o(1)}  G_s(n,t_n^{+},P_n) $, so Lemma \ref{cp_urig_brig} is now
established.

\end{document}